\documentclass[journal=nalefd,manuscript=article]{achemso}

\usepackage[version=3]{mhchem} 

\usepackage{color}
\usepackage{subfigure}
\usepackage{multirow}

\usepackage{latexsym}
\usepackage{dcolumn}

\newcommand{\beq}{\begin{equation}}
\newcommand{\beqa}{\begin{eqnarray}}
\newcommand{\eeq}{\end{equation}}
\newcommand{\eeqa}{\end{eqnarray}}
\def \bmath #1 {{\hbox{\boldmath{$#1$}\unboldmath}}}

\title[]{\bf Electron transport signature of H$_2$ dissociation 
on atomic gold wires
     }

\author{Alexandre Zanchet} 
\affiliation{Unidad Asociada UAM-CSIC,
  Instituto de F{\'\i}sica Fundamental,
  CSIC Serrano 123, 28006 Madrid, Spain.}
\author{Ana{\' \i}s Dorta-Urra}
\affiliation{Unidad Asociada UAM-CSIC,
  Instituto de F{\'\i}sica Fundamental,
  CSIC Serrano 123, 28006 Madrid, Spain.}
\author{Octavio Roncero}
\affiliation{Unidad Asociada UAM-CSIC,
  Instituto de F{\'\i}sica Fundamental,
  CSIC Serrano 123, 28006 Madrid, Spain.}
\author{Alfredo Aguado}
\affiliation{Unidad Asociada UAM-CSIC,
  Departamento de Qu{\'\i}mica F{\'\i}sica,
  Facultad de Ciencias C--XIV,
  Universidad Aut{\'o}noma de Madrid, 28049 Madrid, Spain.}
\author{Jos\'e Ignacio Mart{\'\i}nez}
     \affiliation{
Dept. de Superficies y Recubrimientos, Instituto de Ciencia de Materiales de Madrid (CSIC), 
ES-28049, Madrid, Spain}
\author{Fernando Flores}
     \affiliation{ Departamento de F{\'\i}sica Te\'orica de la Materia
       Condensada,
       Facultad de Ciencias C-V,
       Universidad Aut\'onoma de Madrid, 28049, Madrid, Spain }
\author{Nicol\' as Lorente }
\email{nicolas.lorente@cin2.cat}
\affiliation{
 ICN2 - Institut Catala de Nanociencia i Nanotecnologia, Campus UAB, 08193 Bellaterra (Barcelona), Spain
 }
\affiliation{
CSIC - Consejo Superior de Investigaciones Cientificas, ICN2 Building, 08193 Bellaterra (Barcelona), Spain
}

\begin{document}

\newpage

\begin{abstract}
Non-equilibrium Green's functions calculations based on density functional
theory show a direct link between the initial stages of H$_2$ dissociation
on a gold atomic wire and the electronic current supported by the
gold wire.  The simulations reveal that for biases below the stability
threshold of the wire, the minimum-energy path for H$_2$ dissociation
is not affected. However, the electronic current presents a dramatic
drop when the molecule initiates its dissociation.  This current drop
is traced back to quantum interference between electron paths when the
molecule starts interacting with the gold wire.  
\end{abstract}



The field of molecular dynamics studied by
electronic currents has been continuously expanding in the last two
decades.~\cite{Eigler,Avouris,Gimzewski,Rieder,Ho,Kim,Thomas,Moeller}
Major progress was achieved, when electron currents were taylored
to detect reactions within a single molecule, prompting the search
for single-molecule chemistry.~\cite{Hla,Komeda,Nacho}
In this way, molecules could be studied to dissociate,~\cite{Avouris}
to assemble,~\cite{Rieder,Ho} to desorb~\cite{Komeda,Nacho} in the
extraordinary conditions of a perfectly known environment. All of these
experiments were perfomed for molecular adsorbates under a scanning
tunneling microscope tip. However, no electrical probe has been proposed
to study single-molecule chemistry in gas phase or in solution.

Using the variations of electronic currents in nanowires is a very
interesting possibility for reaching the single-molecule limit in
gas or liquid chemical reactions.  Due to the extreme reduction in
the wire's dimensions, nanowires present enhanced reactivity. Thus,
the very inert noble metal, gold, becomes chemically active in
clusters~\cite{Hutchings:85,Haruta-etal:89,Sanchez-etal:99,Hakkinen-Landman:00,Hakkinen-etal:02,Gilb-etal:02,%
Furche-etal:02,Fernandez-etal:04,Gruene-etal:08,Lechtken-etal:09}
and nanowires~\cite{Yanson-etal:98,Ohnishi-etal:98,Hakkinen-etal:99,%
Bahn-etal:02,Legoas-etal:02,Csonka-etal:03,Barnett-etal:04,Frederiksen-etal:07}
due to the presence of many dangling orbitals which
form bonds with surrounding molecules. In particular, the possible dissociation
 of H$_2$ on gold nanowires has already been 
reported\cite{Jelinek-etal:06,Zanchet-etal:09a}.  In order
to use the nanowire as a reactivity sensor, the electronic transport
properties of the wire should change in presence of a chemical reaction
while the reaction remains unaltered by the flowing electron current.
Unfortunately, no evidence of the reactivity-sensing properties of
atom-size nanowires has been revealed yet.

%
%
%
%
{In this Letter, we concurrently evaluate the minimum-energy path (MEP)
of a gas-phase H$_2$ molecule that impinges on a gold nanowire
and the wire's electron current, using
non-equilibrium Green's functions (NEGF) and density functional theory (DFT).}
These calculations show that for biases below the
stability threshold of the wire, the H$_2$ molecular dynamics is not
affected. However, the electronic current presents a dramatic drop when
the molecule initiates its dissociation. Hence, atom-size nanowires
can be excellent probes for determining the onset of H$_2$ dissociation
without perturbation from the measuring current.


Starting from an initial atomic configuration, the MEP is obtained by
computing the total potential energy and its gradient as the H$_2$
molecule approaches the atomic wire.  The initial configuration
corresponds to H$_2$ far from a gold monoatomic wire and the final
one corresponds to the chemisorbed species on the wire, forming the
usual Au-H-Au-H-Au double-bridge bond.~\cite{Zanchet-etal:09a} The energy
and corresponding gradients  are evaluated using the TRANSIESTA
code~\cite{Soler-etal:02,Brandbyge-etal:02} with the Perdew, Burke and
Ernzerhof (PBE96) functional.~\cite{Perdew-etal:96}
The whole system can be divided in three distinct regions breaking
periodicity along the transport direction.  The contact region is a
central region composed of a 4-atom free-standing gold chain connected
to two electrodes, left and right, each formed by the first 5 layers
of a $3\times3$ cell of Au (111).  The two electrodes were frozen,
and kept at a distance of 13.3 \AA. The inter-electrode distance
affects the inter-atomic distance, which has been 
analyzed elsewhere.~\cite{Barnett-etal:04}  The other two regions are the
semi-infinite electrodes formed by periodically repeating 3 layers
of bulk gold in the (111) direction.  In this way, bias can be applied
between the two asymptotic regions by shifting their Fermi levels, and by
self-consistently solving the Poisson and DFT equations
for the non-equilibrium density matrix in the central
region.~\cite{Brandbyge-etal:02} Finally, the electron transmission
through the central region is evaluated for all cases using
standard non-equilibrium Green's functions.~\cite{Brandbyge-etal:02}



\begin{figure}[t]
\centerline{\includegraphics[width=0.6\columnwidth]{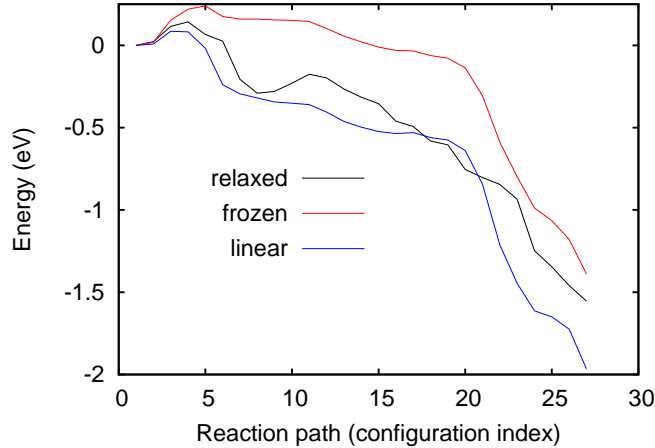}}
\caption{(Color online) {Minimum-energy path (MEP) for three models of a hydrogen molecule
impinging on a 4-atom gold wire connected to semi-infinite electrodes.
The energy  with respect to the initial configuration energy is plotted against the index
of the sequence of minimum-energy configurations of the H$_2$-wire system. The initial
configuration corresponds to relaxed molecule and wire distant enough to have no mutual interaction.
Black line (relaxed): MEP when the 4-atom chain is also allowed to change its geometry.
Red line (frozen): MEP for a frozen 4-atom chain to the initial configuration.
Blue line (linear): MEP for an infinite gold atom chain that is allowed to adapt to the
evolving  H$_2$  molecule. The realistic case (relaxed) shows two equilibrium situations
at the stages $i$=8 and $i$=27 of the MEP.}
}
\label{mepV0}
\end{figure}

Three different  models of the wire, of decreasing complexity, have been
built to better understand the processes under study, and in each case
a MEP has been calculated. First, the model contact described above
in which the 4 central gold and the two hydrogen  atoms are optimized
along the MEP (relaxed wire).  In the second case, the 4 central gold
atoms are kept frozen in a linear configuration, and only the geometry
of the two hydrogen atoms are optimized (frozen wire).  Finally, we
consider the case of a linear chain of 12 gold atoms (linear wire),
repeated periodically, kept frozen and where only the two hydrogen
atoms define the MEP.  The MEP in the three considered cases have been
discretized in a set of 27 points, each one corresponding to a different
nuclear configuration. The initial stage, $i$=1, corresponds to the H$_2$ far from the wire,
while $i=27$ corresponds to the hydrogen atoms chemisorbed on the wire
in the Au-H-Au-H-Au double-bridge bond.  In between these two
limiting configurations, the label $i$ denotes similar geometries for the three
MEPs considered as shown in ~\ref{mepV0}

\begin{figure}[h]
\centerline{\includegraphics[width=\columnwidth]{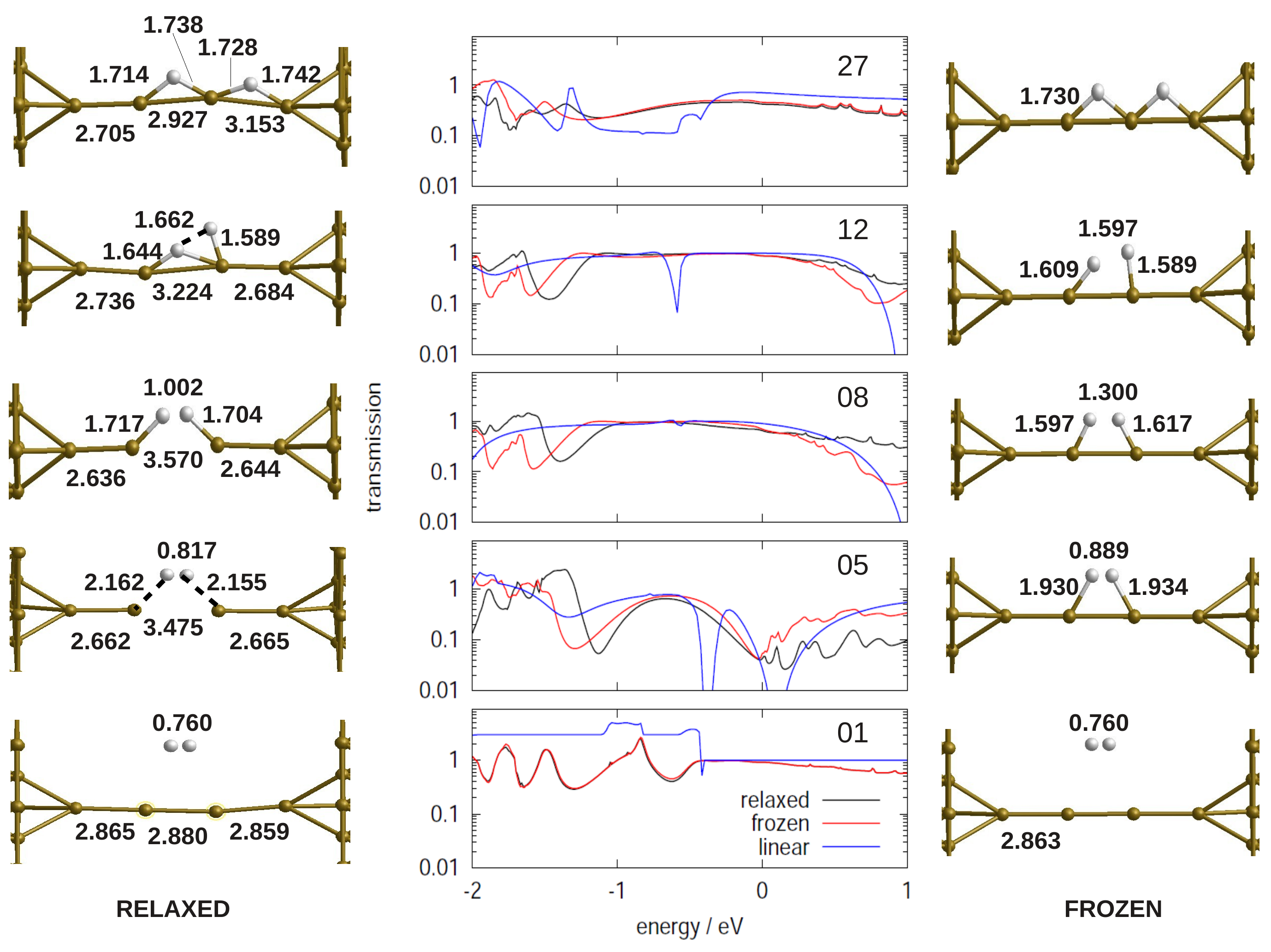}}
\caption{(Color online) {Central panels: electronic transmission versus electronic energy at 
significant points along the H$_2$  MEP. The zero of energy is taken at the electrode's Fermi level.
In the lateral panels, the geometries of the relaxed (left) and frozen (right) wires are shown.
The geometries of the infinite linear gold wire are very similar to those of the frozen wire.
Numbers in the geometrical schemes denote
distances in~\AA. The spontaneous H$_2$ dissociation is initiated at stage $i$=5, where the
transmission shows an important decrease at the Fermi level.}}
\label{geometry-transmision}
\end{figure}

In the three cases, H$_2$ dissociation on the gold wire is exothermic by
1.4-2 eV. All the cases present a low barrier when H$_2$ approaches the
wire (approximately at $i$=4). The barrier height is $\approx$ 0.12 eV
for the relaxed and linear wires and $\approx$ 0.25 eV for the frozen
case. Keeping in mind that the zero-point energy of H$_2$ is $\approx$
0.25 eV, it should be easy to overpass. Once
on the top of this barrier, the reaction path has been determined by
following the gradient.
In the case of the relaxed wire, there is a well for $i$=8, of roughly
0.25 eV below the asymptote given by $i$=1. 
Some
characteristic geometries for the relaxed and frozen wires are illustrated
in the lateral panels of ~\ref{geometry-transmision}

We can now evaluate the electron transmission
between electrodes for each of the stages, $i$, of the MEP. When H$_2$ is
far from the wire, $i=1$ bottom panel of \ref{geometry-transmision},
the transmission at the Fermi level is equal to one for the three cases
because each system presents one single conduction channel.  On the
opposite side, for $i$=27 with the two hydrogen atoms chemisorbed in
the wire forming the double-bridge bond, the transmission is reduced,
being $\approx$ 0.75 in the linear model and  $\approx$ 0.5 in the
other two cases.  At the minimum ($i$=8) and saddle ($i$=12) points
the transmission is in between 0.8 and 1 in the three cases, all
of them showing a rather smooth and similar behavior along the MEP.
However, $i$=5 is a striking point because the three cases show a nearly
zero transmission. It corresponds to the initial stage of the H$_2$
dissociation where a slightly elongated H$_2$ starts to attach to the
gold wire.

\begin{figure}[t]
\centerline{\includegraphics[width=0.8\columnwidth]{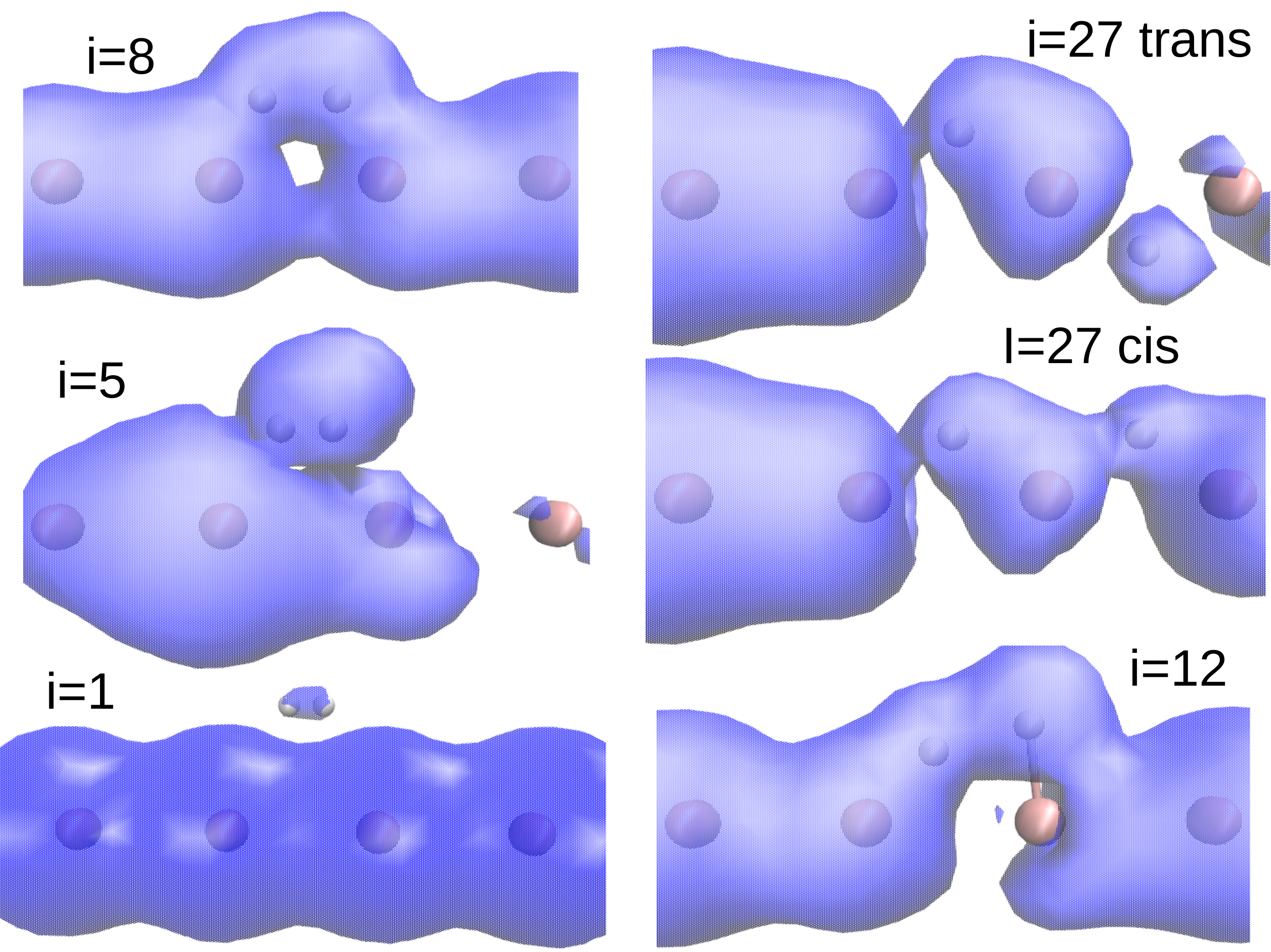}}
\caption{(Color online) Eigenchannels obtained at different $i$ points along the MEP,
 for the case of the linear wire. For the MEP stage $i$=5, the electronic eigenchannel is
interrupted just passed the molecule, indicating a drop in electron flux.}
\label{eigenchannels}
\end{figure}

To analyze this result, and based on
the similarity of the three model systems considered here,
we use the simplest linear model.  We have calculated its
eigenchannels, defined as the non-mixing transmission channels
that diagonalize the transmission matrix with eigenvalues $0\leq
T_n \leq 1$, whose sum provides the total transmission of the
system.~\cite{Mujica-etal:94,Kopf-Saalfrank:04,Paulsson-Brandbyge:07}  For
all the geometries, there is only one conduction eigenchannel contributing
to the transmission at the Fermi energy, as expected for gold
atomic chains. The square of the
modulus of these eigenchannels provide a visualization of the electron
flux though the wire, and it is plotted in \ref{eigenchannels}.

Clearly, when H$_2$ is far from the wire ($i$=1) the conduction channel
is located in the wire, \ref{eigenchannels} bottom left. When
H$_2$ sticks to the wire, part of the electron flux passes through it,  
either as molecular hydrogen ($i=8$) or as atomic hydrogen
chemisorbed on the wire ($i=$27).  At $i$=5, however, the amplitude
density of the eigenchannel becomes zero at the right edge, indicating
that the electron flux is stopped. {This reveals an interference effect.}

\begin{figure}[t]
\centerline{\includegraphics[width=0.8\columnwidth]{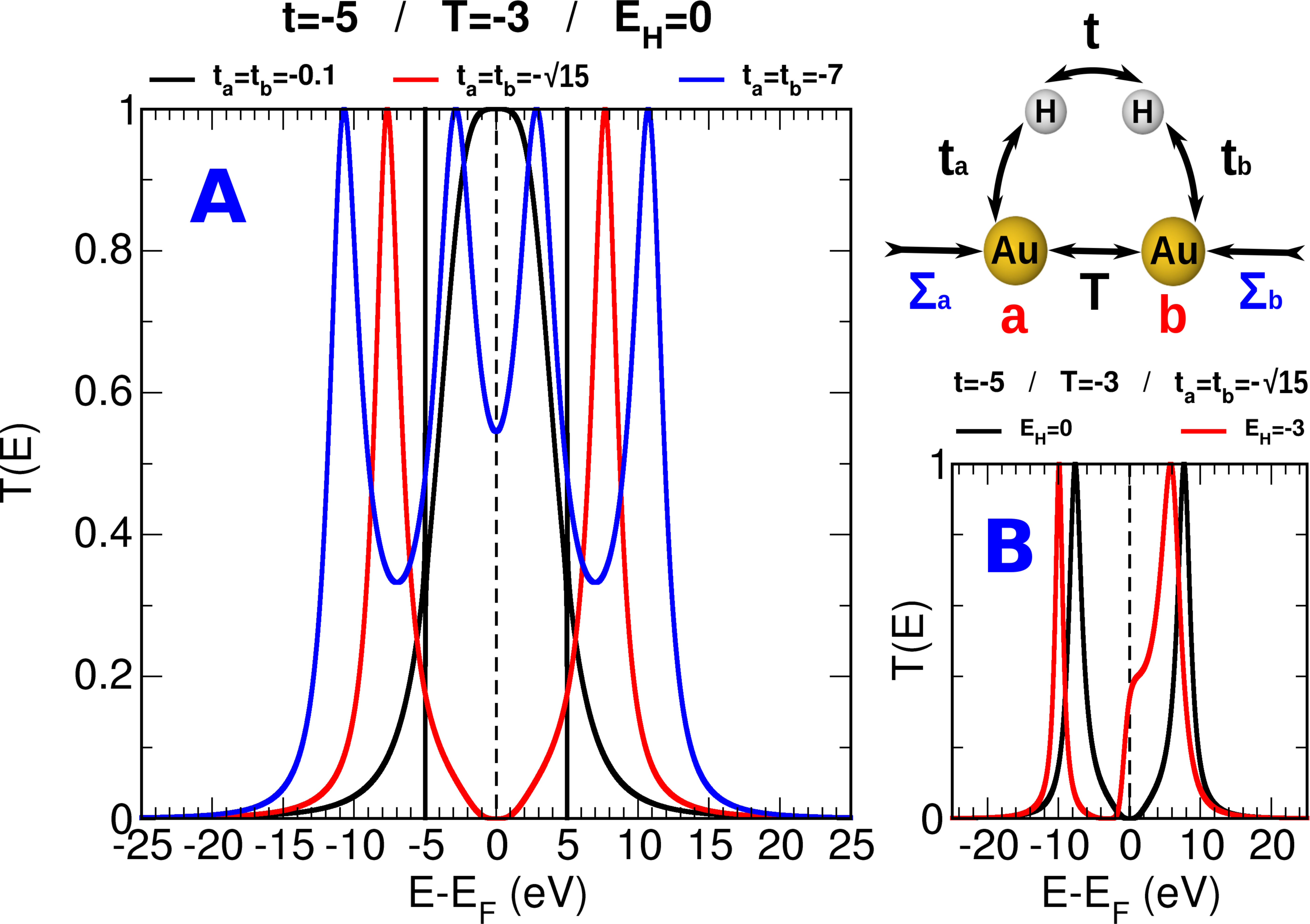}}
\caption{(Color online) {
 Transmission as a function of energy for the model system shown in the upper 
right-hand side of the figure, for different values of $t_a=t_b$ 
and $E_H$ ($E_{Au}$=0 eV, $T$=-3 eV and t=-5eV).
 Panel A: $E_H$=0. Panel B: $E_H$=-3 eV. For $t_a=t_b=\sqrt{15}, t_a^2=Tt$  a dip in $T(E)$ is found.}}
\label{model}
\end{figure}

This interference behavior can be rationalized in terms of a 4-state tight-binding
model composed by 2 gold and 2 hydrogen atoms described by one orbital, where the possible
electronic paths along the wire and the molecule are jointly solved. The model is defined
by the energies $E_{Au}$=0 and $E_H$, and by three hopping terms $T$, $t_a=t_b$ and $t$ (see \ref{model}).
The coupling to the rest of the system is characterized by the self-energies $\Sigma_a=\Sigma_b$
associated with the a or b leads; then the wire's transmission, $T(E)$, is given by\cite{Fisher-Lee:81}
\begin{eqnarray}
T(E)={8 e^2\over h} \left\vert \Sigma_a(E)\right\vert^2 \left\vert G_{ab}(E)\right\vert^2,
\end{eqnarray}
where $G_{ab}(E)$ is the $ab$-component of the Green-function associated with the four-atom model described above.
~\ref{model} shows $T(E)$ for the indicated set of parameters, taking $\Sigma_a=\Sigma_b=iT$ to
guarantee that $T(E=0)$ becomes 1, for $t_a=t_b=0$, the case of an ideal Au-chain,
as in the point $i$=1. As H$_2$ approaches the wire, $t_a$ grows and the electron can also 
be transmitted through
the Au-H bond with two concurrent electron paths competing, one going through the H-atoms
 and the other one through
the Au-atoms. There is a critical situation, when $t_a^2=Tt$, in which the two conducting paths interfere
destructively bringing the transmission to zero (see~\ref{model}).
\cite{solomon_exploring_2010,guedon_observation_2012,Markussen-etal:11}.
 This
interference is the analogue to the one manifested as a Fano profile in
photoionization~\cite{Fano:61}, when the absorption cross section presents a minimum at
energies in which the direct photon excitation to a dissociative continuum
(electron conduction through the gold atoms) interferes with the indirect
mechanism, in which a quasi-bound state is first reached  by the photon
(electron localization on H atoms), and then predissociates towards
the same dissociative continuum.

These conditions for the interference are imposed
by the atomic geometry found in the MEP. The connection between
geometry and quantum interference in molecular conduction
has already been proven experimentally.~\cite{Venkataraman2012}.
In the present case,
the H$_2$-wire interactions are strong enough to
steer the molecule to become parallel to the wire before
dissociation. In this geometry (point $i=5$ of the MEP), the
electronic conditions needed for the above interference are met.

When studying the frozen and the relaxed nano-wires, 
periodic conditions on $(x,y)$ are considered in order to
build the electrodes, this permits us to define
parallel $k$-vectors. For each $(k_x,k_y)$ there is a similar pattern to the one
discussed above for the linear case, but the position
of the resonance varies with $(k_x,k_y)$. Thus, when averaging to get the total
transmission, the zero at $i$=5 is washed out but the transmission dumps
to 0.05 
at the Fermi level, considerably
lower than for the rest of the MEP points,
~\ref{geometry-transmision}. This indicates that the
sudden decrease of the transmission persists in realistic gold wires
when a molecule like H$_2$ approaches it.



\begin{figure}[t]
\centerline{\includegraphics[width=0.6\columnwidth]{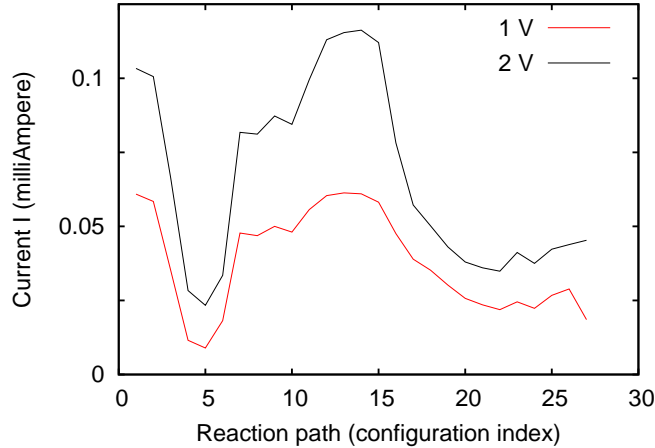}}
\caption{(Color online) {Electron current, $I$, for biases of 1 V and 2 V, \textit{versus}
the system configurations,
which correspond to the zero-bias MEP for the relaxed wire, ~\ref{mepV0}. 
At the dissociation onset, stage $i$=5, there is an important drop in the current due
to quantum interference between the through-wire and through-molecule electron paths.
The drop is visible at both biases. At 2 V, the  wire's stability is compromised and
the zero-bias MEP is used for illustration of the robustness of the
quantum interference effect.
 }}
\label{current-voltage}
\end{figure}

{Finite bias changes the above picture only
quantitatively. The electronic transmission has been investigated at
0.5 V, 1 V and 2 V.  This has been done at the configurations of the
MEP calculated for zero voltage. At finite bias, the MEP is ill-defined
due to the appearance of non-conservative forces.~\cite{dundas_current-driven_2009,Todorov2010,lu_blowing_2010}
However, if the bias-induced changes in the forces are small enough, we expect the zero-voltage MEP
to be an excellent approximation to the molecular path at low temperatures.
Indeed, this is the case.
In order to show this, we studied the minima,
points $i$=8 and 27 of ~\ref{mepV0}. These points correspond
to equilibrium geometries and are then very sensitive to variations
of the forces.  The  bias-induced
forces were negligible and the equilibrium geometries analyzed were the same
as for zero bias. 

For 2 V, the wire is reaching its
stability limit and it is expected to break.~\cite{Yasuda-Sakai:97}
In this case, the system undergoes important changes diverging from
the zero-voltage MEP. For $i$=8
the two hydrogen atoms trend to insert themselves in the gold wire, producing a separation
of the two closest gold atoms. For $i$=27, one hydrogen atom moves to
the other side of the wire, towards a \textit{trans} configuration. 
Hence, the electron-transmissions evaluated for the zero-voltage MEP, while correct at 1 V, are just of
academic interest at 2 V.

\ref{current-voltage} shows the wire's currents evaluated from
the electron transmissions for 1 and 2 V.
As voltage increases, the current increases as well. For all the 
considered wire models, there is a pronounced decrease of the current intensity for
$i$=5. Again, we retrieve that the
decrease is due to the quantum-interference effect at the initial stage
of dissociation. These results show that even at biases larger than 1 V, within
the stability limit of the wire, the dissociation of the $H_2$ molecule
will cause a measurable decrease in the electron current through
the gold wire.}

In summary, non-equilibrium Green's functions (NEGF)
calculations based on density functional theory (DFT) are used to evaluate
the minimal energy path (MEP) that an H$_2$ molecule follows near an
atomic-sized gold wire. The molecule dissociates on the wire and the
electronic current decrease in the initial stages of H$_2$ dissociation
on a gold atomic wire.  This current drop is due to quantum interference
between electron paths when the molecule starts interacting with the
gold wire. Moreover, no significant dependence of the MEP on applied
biases below the wire's stability limit has been found. 
This suggests that the conductance properties of
gold wires can be recorded to acquire information on chemical reactions
taken place in the single-molecule limit in a gas or liquid environment.

This work has been supported by Comunidad Aut\'onoma de Madrid (CAM) under
Grant No. S-2009/MAT/1467, by the  Ministerio de Ciencia e Innovaci\'on
under Grant No. FIS2011-29596-C02, and by the European-Union Integrated
Project AtMol (http://www.atmol.eu). We would like to thank as well
the CESGA computing centre for the computing time under the ICTS grants.

\bibliography{Zanchet}

\providecommand*{\mcitethebibliography}{\thebibliography}
\csname @ifundefined\endcsname{endmcitethebibliography}
{\let\endmcitethebibliography\endthebibliography}{}
\begin{mcitethebibliography}{48}
\providecommand*{\natexlab}[1]{#1}
\providecommand*{\mciteSetBstSublistMode}[1]{}
\providecommand*{\mciteSetBstMaxWidthForm}[2]{}
\providecommand*{\mciteBstWouldAddEndPuncttrue}
  {\def\EndOfBibitem{\unskip.}}
\providecommand*{\mciteBstWouldAddEndPunctfalse}
  {\let\EndOfBibitem\relax}
\providecommand*{\mciteSetBstMidEndSepPunct}[3]{}
\providecommand*{\mciteSetBstSublistLabelBeginEnd}[3]{}
\providecommand*{\EndOfBibitem}{}
\mciteSetBstSublistMode{f}
\mciteSetBstMaxWidthForm{subitem}{(\alph{mcitesubitemcount})}
\mciteSetBstSublistLabelBeginEnd{\mcitemaxwidthsubitemform\space}
{\relax}{\relax}

\bibitem[Eigler et~al.(1991)Eigler, Lutz, and Rudge]{Eigler}
Eigler,~D.; Lutz,~C.; Rudge,~W. \emph{Nature} \textbf{1991}, \emph{352},
  600--603\relax
\mciteBstWouldAddEndPuncttrue
\mciteSetBstMidEndSepPunct{\mcitedefaultmidpunct}
{\mcitedefaultendpunct}{\mcitedefaultseppunct}\relax
\EndOfBibitem
\bibitem[Dujardin et~al.(1992)Dujardin, Walkup, and Avouris]{Avouris}
Dujardin,~G.; Walkup,~R.; Avouris,~P. \emph{Science} \textbf{1992}, \emph{255},
  1232--1235\relax
\mciteBstWouldAddEndPuncttrue
\mciteSetBstMidEndSepPunct{\mcitedefaultmidpunct}
{\mcitedefaultendpunct}{\mcitedefaultseppunct}\relax
\EndOfBibitem
\bibitem[Gimzewski et~al.(1998)Gimzewski, Joachim, Schlittler, Langlais, Tang,
  and Johannsen]{Gimzewski}
Gimzewski,~J.; Joachim,~C.; Schlittler,~R.; Langlais,~V.; Tang,~H.;
  Johannsen,~I. \emph{Science} \textbf{1998}, \emph{281}, 531--533\relax
\mciteBstWouldAddEndPuncttrue
\mciteSetBstMidEndSepPunct{\mcitedefaultmidpunct}
{\mcitedefaultendpunct}{\mcitedefaultseppunct}\relax
\EndOfBibitem
\bibitem[Hla et~al.(2000)Hla, Bartels, Meyer, and Rieder]{Rieder}
Hla,~S.-W.; Bartels,~L.; Meyer,~G.; Rieder,~K.-H. \emph{Phys. Rev. Lett.}
  \textbf{2000}, \emph{85}, 2777--2780\relax
\mciteBstWouldAddEndPuncttrue
\mciteSetBstMidEndSepPunct{\mcitedefaultmidpunct}
{\mcitedefaultendpunct}{\mcitedefaultseppunct}\relax
\EndOfBibitem
\bibitem[Lee and Ho(1999)]{Ho}
Lee,~H.; Ho,~W. \emph{Science} \textbf{1999}, \emph{286}, 1719--1722\relax
\mciteBstWouldAddEndPuncttrue
\mciteSetBstMidEndSepPunct{\mcitedefaultmidpunct}
{\mcitedefaultendpunct}{\mcitedefaultseppunct}\relax
\EndOfBibitem
\bibitem[Sainoo et~al.(2005)Sainoo, Kim, Okawa, Komeda, Shigekawa, and
  Kawai]{Kim}
Sainoo,~Y.; Kim,~Y.; Okawa,~T.; Komeda,~T.; Shigekawa,~H.; Kawai,~M.
  \emph{Phys. Rev. Lett.} \textbf{2005}, \emph{95}, 246102\relax
\mciteBstWouldAddEndPuncttrue
\mciteSetBstMidEndSepPunct{\mcitedefaultmidpunct}
{\mcitedefaultendpunct}{\mcitedefaultseppunct}\relax
\EndOfBibitem
\bibitem[Kumagai et~al.(2012)Kumagai, Shiotari, Okuyama, Hatta, Aruga, Hamada,
  Frederiksen, and Ueba]{Thomas}
Kumagai,~T.; Shiotari,~A.; Okuyama,~H.; Hatta,~S.; Aruga,~T.; Hamada,~I.;
  Frederiksen,~T.; Ueba,~H. \emph{Nature Materials} \textbf{2012}, \emph{11},
  167--172\relax
\mciteBstWouldAddEndPuncttrue
\mciteSetBstMidEndSepPunct{\mcitedefaultmidpunct}
{\mcitedefaultendpunct}{\mcitedefaultseppunct}\relax
\EndOfBibitem
\bibitem[Schaffert et~al.(2013)Schaffert, Cottin, Sonntag, Karacuban, Bobisch,
  Lorente, Gauyacq, and Moeller]{Moeller}
Schaffert,~J.; Cottin,~M.~C.; Sonntag,~A.; Karacuban,~H.; Bobisch,~C.~A.;
  Lorente,~N.; Gauyacq,~J.-P.; Moeller,~R. \emph{Nature Materials}
  \textbf{2013}, \emph{12}, 223--227\relax
\mciteBstWouldAddEndPuncttrue
\mciteSetBstMidEndSepPunct{\mcitedefaultmidpunct}
{\mcitedefaultendpunct}{\mcitedefaultseppunct}\relax
\EndOfBibitem
\bibitem[Hla et~al.(2004)Hla, Braun, Wassermann, and Rieder]{Hla}
Hla,~S.~W.; Braun,~K.~F.; Wassermann,~B.; Rieder,~K.~H. \emph{Physical Review
  letters} \textbf{2004}, \emph{93}, 208302\relax
\mciteBstWouldAddEndPuncttrue
\mciteSetBstMidEndSepPunct{\mcitedefaultmidpunct}
{\mcitedefaultendpunct}{\mcitedefaultseppunct}\relax
\EndOfBibitem
\bibitem[Komeda et~al.(2002)Komeda, Kim, Kawai, Persson, and Ueba]{Komeda}
Komeda,~T.; Kim,~Y.; Kawai,~M.; Persson,~B.; Ueba,~H. \emph{Science}
  \textbf{2002}, \emph{295}, 2055--2058\relax
\mciteBstWouldAddEndPuncttrue
\mciteSetBstMidEndSepPunct{\mcitedefaultmidpunct}
{\mcitedefaultendpunct}{\mcitedefaultseppunct}\relax
\EndOfBibitem
\bibitem[Pascual et~al.(2003)Pascual, Lorente, Song, Conrad, and Rust]{Nacho}
Pascual,~J.~I.; Lorente,~N.; Song,~Z.; Conrad,~H.; Rust,~H.~P. \emph{Nature}
  \textbf{2003}, \emph{423}, 525--528\relax
\mciteBstWouldAddEndPuncttrue
\mciteSetBstMidEndSepPunct{\mcitedefaultmidpunct}
{\mcitedefaultendpunct}{\mcitedefaultseppunct}\relax
\EndOfBibitem
\bibitem[Hutchings(1985)]{Hutchings:85}
Hutchings,~G.~J. \emph{J. Catal.} \textbf{1985}, \emph{96}, 292\relax
\mciteBstWouldAddEndPuncttrue
\mciteSetBstMidEndSepPunct{\mcitedefaultmidpunct}
{\mcitedefaultendpunct}{\mcitedefaultseppunct}\relax
\EndOfBibitem
\bibitem[Haruta et~al.(1989)Haruta, Yamada, Kobayashi, and
  Iijima]{Haruta-etal:89}
Haruta,~M.; Yamada,~N.; Kobayashi,~T.; Iijima,~S. \emph{J. Catal.}
  \textbf{1989}, \emph{115}, 301\relax
\mciteBstWouldAddEndPuncttrue
\mciteSetBstMidEndSepPunct{\mcitedefaultmidpunct}
{\mcitedefaultendpunct}{\mcitedefaultseppunct}\relax
\EndOfBibitem
\bibitem[Sanchez et~al.(1999)Sanchez, Abbet, Heiz, W.-D.~Schneider, N, Barnett,
  and Landman]{Sanchez-etal:99}
Sanchez,~A.; Abbet,~S.; Heiz,~U.; W.-D.~Schneider,~H.~H.; N,~R.; Barnett,;
  Landman,~U. \emph{J. Phys. Chem. A} \textbf{1999}, \emph{103}, 9573\relax
\mciteBstWouldAddEndPuncttrue
\mciteSetBstMidEndSepPunct{\mcitedefaultmidpunct}
{\mcitedefaultendpunct}{\mcitedefaultseppunct}\relax
\EndOfBibitem
\bibitem[H\"akkinen and Landman(2000)]{Hakkinen-Landman:00}
H\"akkinen,~H.; Landman,~U. \emph{Phys. Rev. B} \textbf{2000}, \emph{62},
  R2287\relax
\mciteBstWouldAddEndPuncttrue
\mciteSetBstMidEndSepPunct{\mcitedefaultmidpunct}
{\mcitedefaultendpunct}{\mcitedefaultseppunct}\relax
\EndOfBibitem
\bibitem[H\"akkinen et~al.(2002)H\"akkinen, Moseler, and
  Landman]{Hakkinen-etal:02}
H\"akkinen,~H.; Moseler,~M.; Landman,~U. \emph{Phys. Rev. Lett.} \textbf{2002},
  \emph{89}, 033401\relax
\mciteBstWouldAddEndPuncttrue
\mciteSetBstMidEndSepPunct{\mcitedefaultmidpunct}
{\mcitedefaultendpunct}{\mcitedefaultseppunct}\relax
\EndOfBibitem
\bibitem[Gilb et~al.(2002)Gilb, Weis, Furche, Ahlrichs, and
  Kappes]{Gilb-etal:02}
Gilb,~S.; Weis,~P.; Furche,~F.; Ahlrichs,~R.; Kappes,~M.~M. \emph{J. Chem.
  Phys.} \textbf{2002}, \emph{116}, 4094\relax
\mciteBstWouldAddEndPuncttrue
\mciteSetBstMidEndSepPunct{\mcitedefaultmidpunct}
{\mcitedefaultendpunct}{\mcitedefaultseppunct}\relax
\EndOfBibitem
\bibitem[Furche et~al.(2002)Furche, Ahlrichs, Weis, jacob, Gilb, Bierweiler,
  and Kappes]{Furche-etal:02}
Furche,~F.; Ahlrichs,~R.; Weis,~P.; jacob,~C.; Gilb,~S.; Bierweiler,~T.;
  Kappes,~M.~M. \emph{J. Chem. Phys.} \textbf{2002}, \emph{117}, 6982\relax
\mciteBstWouldAddEndPuncttrue
\mciteSetBstMidEndSepPunct{\mcitedefaultmidpunct}
{\mcitedefaultendpunct}{\mcitedefaultseppunct}\relax
\EndOfBibitem
\bibitem[Fern{\'a}ndez et~al.(2004)Fern{\'a}ndez, Soler, Garz{\'o}n, and
  Balb{\'a}s]{Fernandez-etal:04}
Fern{\'a}ndez,~E.~M.; Soler,~J.~M.; Garz{\'o}n,~I.~L.; Balb{\'a}s,~L.~C.
  \emph{Phys. Rev. B} \textbf{2004}, \emph{70}, 165403\relax
\mciteBstWouldAddEndPuncttrue
\mciteSetBstMidEndSepPunct{\mcitedefaultmidpunct}
{\mcitedefaultendpunct}{\mcitedefaultseppunct}\relax
\EndOfBibitem
\bibitem[Gruene et~al.(2008)Gruene, Rayner, Redlich, van~der Meer, Lyon,
  Maijer, and Fielicke]{Gruene-etal:08}
Gruene,~P.; Rayner,~D.~M.; Redlich,~B.; van~der Meer,~A. F.~G.; Lyon,~J.~T.;
  Maijer,~G.; Fielicke,~A. \emph{Science} \textbf{2008}, \emph{321}, 674\relax
\mciteBstWouldAddEndPuncttrue
\mciteSetBstMidEndSepPunct{\mcitedefaultmidpunct}
{\mcitedefaultendpunct}{\mcitedefaultseppunct}\relax
\EndOfBibitem
\bibitem[Lechtken et~al.(2009)Lechtken, Neiss, Kappes, and
  Schooss]{Lechtken-etal:09}
Lechtken,~A.; Neiss,~C.; Kappes,~M.~M.; Schooss,~D. \emph{Phys. Chem. Chem.
  Phys.} \textbf{2009}, \emph{11}, 4344\relax
\mciteBstWouldAddEndPuncttrue
\mciteSetBstMidEndSepPunct{\mcitedefaultmidpunct}
{\mcitedefaultendpunct}{\mcitedefaultseppunct}\relax
\EndOfBibitem
\bibitem[Yanson et~al.(1998)Yanson, Bollinger, van~den Brom, Agrait, and van
  Ruitenbeek]{Yanson-etal:98}
Yanson,~A.~I.; Bollinger,~G.~R.; van~den Brom,~H.~E.; Agrait,~N.; van
  Ruitenbeek,~J.~M. \emph{Nature} \textbf{1998}, \emph{395}, 783\relax
\mciteBstWouldAddEndPuncttrue
\mciteSetBstMidEndSepPunct{\mcitedefaultmidpunct}
{\mcitedefaultendpunct}{\mcitedefaultseppunct}\relax
\EndOfBibitem
\bibitem[Ohnishi et~al.(1998)Ohnishi, Kondo, and Takayanagi]{Ohnishi-etal:98}
Ohnishi,~H.; Kondo,~Y.; Takayanagi,~K. \emph{Nature} \textbf{1998}, \emph{395},
  780\relax
\mciteBstWouldAddEndPuncttrue
\mciteSetBstMidEndSepPunct{\mcitedefaultmidpunct}
{\mcitedefaultendpunct}{\mcitedefaultseppunct}\relax
\EndOfBibitem
\bibitem[H\"akkinen et~al.(1999)H\"akkinen, Barnett, and
  Landman]{Hakkinen-etal:99}
H\"akkinen,~H.; Barnett,~R.~N.; Landman,~U. \emph{J. Phys. Chem. B}
  \textbf{1999}, \emph{103}, 8814\relax
\mciteBstWouldAddEndPuncttrue
\mciteSetBstMidEndSepPunct{\mcitedefaultmidpunct}
{\mcitedefaultendpunct}{\mcitedefaultseppunct}\relax
\EndOfBibitem
\bibitem[Bahn et~al.(2002)Bahn, Lopez, Norskov, and Jacobsen]{Bahn-etal:02}
Bahn,~S.~R.; Lopez,~N.; Norskov,~J.~K.; Jacobsen,~K.~W. \emph{Phys. Rev. B}
  \textbf{2002}, \emph{66}, 081405(R)\relax
\mciteBstWouldAddEndPuncttrue
\mciteSetBstMidEndSepPunct{\mcitedefaultmidpunct}
{\mcitedefaultendpunct}{\mcitedefaultseppunct}\relax
\EndOfBibitem
\bibitem[Legoas et~al.(2002)Legoas, Galvao, Rodrigues, and
  Ugarte]{Legoas-etal:02}
Legoas,~S.~B.; Galvao,~D.~S.; Rodrigues,~V.; Ugarte,~D. \emph{Phys. Rev. lett.}
  \textbf{2002}, \emph{88}, 076105\relax
\mciteBstWouldAddEndPuncttrue
\mciteSetBstMidEndSepPunct{\mcitedefaultmidpunct}
{\mcitedefaultendpunct}{\mcitedefaultseppunct}\relax
\EndOfBibitem
\bibitem[Csonka et~al.(2003)Csonka, Halbritter, Mih\'aly, E.~Jurdik, Speller,
  and van Kempen]{Csonka-etal:03}
Csonka,~S.; Halbritter,~A.; Mih\'aly,~G.; E.~Jurdik,~O.~S.; Speller,~S.; van
  Kempen,~H. \emph{Phys. Rev. Lett.} \textbf{2003}, \emph{90}, 116803\relax
\mciteBstWouldAddEndPuncttrue
\mciteSetBstMidEndSepPunct{\mcitedefaultmidpunct}
{\mcitedefaultendpunct}{\mcitedefaultseppunct}\relax
\EndOfBibitem
\bibitem[Barnett et~al.(2004)Barnett, H\"akkinen, Scherbakov, and
  Landman]{Barnett-etal:04}
Barnett,~R.~N.; H\"akkinen,~H.; Scherbakov,~A.~G.; Landman,~U. \emph{Nano
  Lett.} \textbf{2004}, \emph{4}, 1845\relax
\mciteBstWouldAddEndPuncttrue
\mciteSetBstMidEndSepPunct{\mcitedefaultmidpunct}
{\mcitedefaultendpunct}{\mcitedefaultseppunct}\relax
\EndOfBibitem
\bibitem[Frederiksen et~al.(2007)Frederiksen, Paulsson, and
  Brandbyge]{Frederiksen-etal:07}
Frederiksen,~T.; Paulsson,~M.; Brandbyge,~M. \emph{J. Phys.} \textbf{2007},
  \emph{61}, 312\relax
\mciteBstWouldAddEndPuncttrue
\mciteSetBstMidEndSepPunct{\mcitedefaultmidpunct}
{\mcitedefaultendpunct}{\mcitedefaultseppunct}\relax
\EndOfBibitem
\bibitem[Jel{\'\i}nek et~al.(2006)Jel{\'\i}nek, P{\'e}rez, Ortega, and
  Flores]{Jelinek-etal:06}
Jel{\'\i}nek,~P.; P{\'e}rez,~R.; Ortega,~J.; Flores,~F. \emph{Phys. Rev. Lett.}
  \textbf{2006}, \emph{96}, 046803\relax
\mciteBstWouldAddEndPuncttrue
\mciteSetBstMidEndSepPunct{\mcitedefaultmidpunct}
{\mcitedefaultendpunct}{\mcitedefaultseppunct}\relax
\EndOfBibitem
\bibitem[Zanchet et~al.(2009)Zanchet, Dorta-Urra, Roncero, Flores, Tablero,
  Panigua, and Aguado]{Zanchet-etal:09a}
Zanchet,~A.; Dorta-Urra,~A.; Roncero,~O.; Flores,~F.; Tablero,~C.; Panigua,~M.;
  Aguado,~A. \emph{Phys. Chem. Chem. Phys.} \textbf{2009}, \emph{11},
  10122\relax
\mciteBstWouldAddEndPuncttrue
\mciteSetBstMidEndSepPunct{\mcitedefaultmidpunct}
{\mcitedefaultendpunct}{\mcitedefaultseppunct}\relax
\EndOfBibitem
\bibitem[Soler et~al.(2002)Soler, Artacho, Gale, Garc{\'\i}a, Junquera,
  Ordej\'on, and S\'anchez-Portal]{Soler-etal:02}
Soler,~J.~M.; Artacho,~E.; Gale,~J.; Garc{\'\i}a,~A.; Junquera,~J.;
  Ordej\'on,~P.; S\'anchez-Portal,~D. \emph{J. Phys.: Condens. Matter}
  \textbf{2002}, \emph{14}, 2745\relax
\mciteBstWouldAddEndPuncttrue
\mciteSetBstMidEndSepPunct{\mcitedefaultmidpunct}
{\mcitedefaultendpunct}{\mcitedefaultseppunct}\relax
\EndOfBibitem
\bibitem[Brandbyge et~al.(2002)Brandbyge, Mozos, Ordej\'on, Taylor, and
  Stokbro]{Brandbyge-etal:02}
Brandbyge,~M.; Mozos,~J.~L.; Ordej\'on,~P.; Taylor,~J.; Stokbro,~K. \emph{Phys.
  Rev.B} \textbf{2002}, \emph{65}, 165401\relax
\mciteBstWouldAddEndPuncttrue
\mciteSetBstMidEndSepPunct{\mcitedefaultmidpunct}
{\mcitedefaultendpunct}{\mcitedefaultseppunct}\relax
\EndOfBibitem
\bibitem[Perdew et~al.(1996)Perdew, Burke, and Ernzerhof]{Perdew-etal:96}
Perdew,~J.~P.; Burke,~K.; Ernzerhof,~M. \emph{Phys. Rev. Lett.} \textbf{1996},
  \emph{77}, 3865\relax
\mciteBstWouldAddEndPuncttrue
\mciteSetBstMidEndSepPunct{\mcitedefaultmidpunct}
{\mcitedefaultendpunct}{\mcitedefaultseppunct}\relax
\EndOfBibitem
\bibitem[Mujica et~al.(1994)Mujica, Kemp, and Ratner]{Mujica-etal:94}
Mujica,~V.; Kemp,~M.; Ratner,~M.~A. \emph{J. Chem. Phys.} \textbf{1994},
  \emph{101}, 6849\relax
\mciteBstWouldAddEndPuncttrue
\mciteSetBstMidEndSepPunct{\mcitedefaultmidpunct}
{\mcitedefaultendpunct}{\mcitedefaultseppunct}\relax
\EndOfBibitem
\bibitem[Kopf and Saalfrank(2004)]{Kopf-Saalfrank:04}
Kopf,~A.; Saalfrank,~P. \emph{Chem. Phys. Lett.} \textbf{2004}, \emph{386},
  17\relax
\mciteBstWouldAddEndPuncttrue
\mciteSetBstMidEndSepPunct{\mcitedefaultmidpunct}
{\mcitedefaultendpunct}{\mcitedefaultseppunct}\relax
\EndOfBibitem
\bibitem[Paulsson and Brandbyge(2007)]{Paulsson-Brandbyge:07}
Paulsson,~M.; Brandbyge,~M. \emph{Phys. Rev. B} \textbf{2007}, \emph{76},
  115117\relax
\mciteBstWouldAddEndPuncttrue
\mciteSetBstMidEndSepPunct{\mcitedefaultmidpunct}
{\mcitedefaultendpunct}{\mcitedefaultseppunct}\relax
\EndOfBibitem
\bibitem[Fisher and Lee(1981)]{Fisher-Lee:81}
Fisher,~D.~S.; Lee,~P.~A. \emph{Phys. Rev. B} \textbf{1981}, \emph{23},
  6851\relax
\mciteBstWouldAddEndPuncttrue
\mciteSetBstMidEndSepPunct{\mcitedefaultmidpunct}
{\mcitedefaultendpunct}{\mcitedefaultseppunct}\relax
\EndOfBibitem
\bibitem[Solomon et~al.(2010)Solomon, Herrmann, Hansen, Mujica, and
  Ratner]{solomon_exploring_2010}
Solomon,~G.~C.; Herrmann,~C.; Hansen,~T.; Mujica,~V.; Ratner,~M.~A.
  \emph{Nature Chemistry} \textbf{2010}, \emph{2}, 223--228\relax
\mciteBstWouldAddEndPuncttrue
\mciteSetBstMidEndSepPunct{\mcitedefaultmidpunct}
{\mcitedefaultendpunct}{\mcitedefaultseppunct}\relax
\EndOfBibitem
\bibitem[Guédon et~al.(2012)Guédon, Valkenier, Markussen, Thygesen, Hummelen,
  and van~der Molen]{guedon_observation_2012}
Guédon,~C.~M.; Valkenier,~H.; Markussen,~T.; Thygesen,~K.~S.; Hummelen,~J.~C.;
  van~der Molen,~S.~J. \emph{Nature Nanotechnology} \textbf{2012}, \emph{7},
  305--309\relax
\mciteBstWouldAddEndPuncttrue
\mciteSetBstMidEndSepPunct{\mcitedefaultmidpunct}
{\mcitedefaultendpunct}{\mcitedefaultseppunct}\relax
\EndOfBibitem
\bibitem[Markussen et~al.(2011)Markussen, Stadler, and
  Thygesen]{Markussen-etal:11}
Markussen,~T.; Stadler,~R.; Thygesen,~K.~S. \emph{Phys. Chem. Chem. Phys.}
  \textbf{2011}, \emph{13}, 14311\relax
\mciteBstWouldAddEndPuncttrue
\mciteSetBstMidEndSepPunct{\mcitedefaultmidpunct}
{\mcitedefaultendpunct}{\mcitedefaultseppunct}\relax
\EndOfBibitem
\bibitem[Fano(1961)]{Fano:61}
Fano,~U. \emph{Phys. Rev.} \textbf{1961}, \emph{124}, 1866\relax
\mciteBstWouldAddEndPuncttrue
\mciteSetBstMidEndSepPunct{\mcitedefaultmidpunct}
{\mcitedefaultendpunct}{\mcitedefaultseppunct}\relax
\EndOfBibitem
\bibitem[Aradhya et~al.(2012)Aradhya, Meisner, Krikorian, Ahn, Parameswaran,
  Steigerwald, Nuckolls, and Venkataraman]{Venkataraman2012}
Aradhya,~S.~V.; Meisner,~J.~S.; Krikorian,~M.; Ahn,~S.; Parameswaran,~R.;
  Steigerwald,~M.~L.; Nuckolls,~C.; Venkataraman,~L. \emph{Nano Letters}
  \textbf{2012}, \emph{12}, 1643--1647\relax
\mciteBstWouldAddEndPuncttrue
\mciteSetBstMidEndSepPunct{\mcitedefaultmidpunct}
{\mcitedefaultendpunct}{\mcitedefaultseppunct}\relax
\EndOfBibitem
\bibitem[Dundas et~al.(2009)Dundas, {McEniry}, and
  Todorov]{dundas_current-driven_2009}
Dundas,~D.; {McEniry},~E.~J.; Todorov,~T.~N. \emph{Nature Nanotechnology}
  \textbf{2009}, \emph{4}, 99--102\relax
\mciteBstWouldAddEndPuncttrue
\mciteSetBstMidEndSepPunct{\mcitedefaultmidpunct}
{\mcitedefaultendpunct}{\mcitedefaultseppunct}\relax
\EndOfBibitem
\bibitem[Todorov et~al.(2010)Todorov, Dundas, and McEniry]{Todorov2010}
Todorov,~T.~N.; Dundas,~D.; McEniry,~E.~J. \emph{Phys. Rev. B} \textbf{2010},
  \emph{81}, 075416\relax
\mciteBstWouldAddEndPuncttrue
\mciteSetBstMidEndSepPunct{\mcitedefaultmidpunct}
{\mcitedefaultendpunct}{\mcitedefaultseppunct}\relax
\EndOfBibitem
\bibitem[{L\"u} et~al.(2010){L\"u}, Brandbyge, and Hedegaard]{lu_blowing_2010}
{L\"u},~J.-T.; Brandbyge,~M.; Hedegaard,~P. \emph{Nano Letters} \textbf{2010},
  \emph{10}, 1657--1663\relax
\mciteBstWouldAddEndPuncttrue
\mciteSetBstMidEndSepPunct{\mcitedefaultmidpunct}
{\mcitedefaultendpunct}{\mcitedefaultseppunct}\relax
\EndOfBibitem
\bibitem[Yasuda and Sakai(1997)]{Yasuda-Sakai:97}
Yasuda,~H.; Sakai,~A. \emph{Phys. Rev. B} \textbf{1997}, \emph{56}, 1069\relax
\mciteBstWouldAddEndPuncttrue
\mciteSetBstMidEndSepPunct{\mcitedefaultmidpunct}
{\mcitedefaultendpunct}{\mcitedefaultseppunct}\relax
\EndOfBibitem
\end{mcitethebibliography}

\end{document}